\documentclass{biometrika}

\usepackage{amsmath}
\usepackage{newtxtext}
\usepackage[subscriptcorrection]{newtxmath}

\graphicspath{{./figures/}}

\usepackage[plain,noend]{algorithm2e}

\makeatletter
\renewcommand{\algocf@captiontext}[2]{#1\algocf@typo. \AlCapFnt{}#2} % text of caption
\def\@algocf@capt@plain{top}
\renewcommand{\algocf@makecaption}[2]{%
  \addtolength{\hsize}{\algomargin}%
  \sbox\@tempboxa{\algocf@captiontext{#1}{#2}}%
  \ifdim\wd\@tempboxa >\hsize%     % if caption is longer than a line
    \hskip .5\algomargin%
    \parbox[t]{\hsize}{\algocf@captiontext{#1}{#2}}% then caption is not centered
  \else%
    \global\@minipagefalse%
    \hbox to\hsize{\box\@tempboxa}% else caption is centered
  \fi%
  \addtolength{\hsize}{-\algomargin}%
}
\makeatother

\def\AIC{\textsc{aic}}
\def\T{{ \mathrm{\scriptscriptstyle T} }}

\usepackage{url}

\usepackage{algorithmic}
\usepackage{multirow}

\newcommand{\pr}{{\mathrm{pr}}}
\newcommand{\mbf}[1]{{#1}}
\newcommand{\mbb}[1]{{#1}}
\newcommand{\X}{{X}}
\newcommand{\bb}{{\beta}}
\newcommand{{\bt}}{{\theta}}

\usepackage{xr}
\usepackage[table]{xcolor}

\usepackage[normalem]{ulem}
\newcommand{\done}[1]{}

\begin{document}

% \jname{Biometrika}
% %% The year, volume, and number are determined on publication
% \jyear{2025}
% \jvol{112}
% \jnum{1}
% \cyear{2025}
% %% The \doi{...} and \accessdate commands are used by the production team
% %\doi{10.1093/biomet/asm023}
% \accessdate{Advance Access publication on 14 February 2025}

% %% These dates are usually set by the production team
% \received{2 January 2024}
% \revised{3 February 2025}

%% The left and right page headers are defined here:
\markboth{Hu \and Gu}{Survival probability transfer for interval-censored data}

%% Here are the title, author names and addresses
\title{SPOT-IC: Improving prediction for interval-censored data via survival probability transfer}

\author{JINGYI HU \and YU GU}
\affil{Department of Statistics and Actuarial Science, The University of Hong Kong,\\ Pokfulam Road, Hong Kong \email{its\_jingyi@connect.hku.hk}\email{yugu@hku.hk}}

\maketitle
\thispagestyle{empty}

\begin{abstract}
Accurate prediction with interval-censored data is particularly challenging when censoring intervals are wide and follow-up is limited, as is common in studies of chronic diseases. Although auxiliary information from source studies may improve prediction in a target study, existing transfer learning methods typically impose restrictive assumptions on model or parameter similarity, or require access to individual-level source data. We propose a novel transfer learning method for interval-censored data that allows arbitrary source models and avoids sharing source data. Our approach transfers survival probability information from source studies through a carefully designed penalty and enables efficient computation via a simple EM algorithm. When multiple source studies are available and their informativeness is unknown, we further develop a data-adaptive aggregation procedure that is robust to negative transfer. Theoretical analysis shows that the proposed estimator attains a faster convergence rate than the target-only estimator whenever at least one source study is sufficiently informative. Extensive simulation studies and an application to data from the Alzheimer's Disease Neuroimaging Initiative demonstrate the effectiveness of our approach.
\end{abstract}


\section{Introduction}\label{sec-intro}
In many studies of chronic diseases, the disease of interest is asymptomatic or lacks definitive symptoms, so disease onset can only be ascertained through periodic examinations. The resulting failure time data are known as interval-censored data, in the sense that the failure time is only known to lie within a time interval. The absence of exact failure times poses great challenges for prediction with interval-censored data, particularly when the study has a small sample size, short follow-up period, or infrequent examinations.

By leveraging auxiliary information from related source studies, transfer learning provides a powerful tool for improving prediction in a target study. A large body of work has been devoted to transfer learning in various statistical problems \citep{cai2021wei,bastani2021predicting,lis2022transfer,tian2023transfer,lis2024estimation,gu2024angle,cai2024fda}. However, relatively little attention has been paid to transfer learning for survival analysis. Most existing methods assume Cox or transformation models for both the target and source studies and employ an $L_q$-penalty ($q\in[0,2]$) on parameter differences to facilitate parameter sharing \citep{li2016transfer,lizy2023accommodating,xie2024transfer,lu2025adaptive,lou2026two}. These methods have several limitations. First, parameter-level transfer requires the target and source studies to share the same model structure and covariate set, which can be restrictive in practice, especially with multiple source studies. Second, the assumption of similar baseline hazard functions across studies is often violated due to heterogeneity in study populations or follow-up schemes. Third, some methods require access to individual-level source data \citep{li2016transfer,lu2025adaptive,lou2026two}, limiting their applicability in settings involving modern biobanks and electronic health records due to privacy concerns. Although \citet{gu2026prediction} addressed these limitations by transferring predictive information rather than model parameters, their method was developed for right-censored data and scales poorly to multi-source settings due to the need for simultaneous tuning of a large number of parameters.

In this article, we develop a novel transfer learning method for interval-censored data by transferring survival probability information from source studies. We call the proposed method SPOT-IC, short for Survival PrObability Transfer for Interval-Censored data. Unlike existing methods that transfer parameter estimates under a specific source model, SPOT-IC transfers survival probability estimates obtained from arbitrary source models and is therefore considerably more flexible in accommodating heterogeneous source data. Moreover, SPOT-IC only requires the target and source studies to share similar survival probabilities within the target study domain, which is a weaker assumption than the parameter-similarity assumptions imposed by most existing methods. Finally, SPOT-IC accommodates multiple source studies through a data-adaptive aggregation strategy, thereby providing protection against negative transfer with much less tuning effort than the method of \citet{gu2026prediction}.

For the target study, SPOT-IC adopts flexible semiparametric transformation models with potentially time-dependent covariates, while leaving the source models completely unrestricted and requiring no sharing of individual-level source data. To facilitate survival probability transfer, we incorporate the cross-entropy-type penalty and develop a simple and stable computational algorithm based on the equivalence between the penalized objective function and a weighted log-likelihood for mixed interval-censored data. When multiple source studies with unknown informativeness are available, we further propose an aggregation procedure that adaptively combines a collection of candidate estimators. We establish rigorous asymptotic theory and show that the proposed estimator achieves a faster convergence rate than the target-only estimator whenever at least one source estimator is sufficiently accurate for the target population. 
% Our theory requires substantial innovations beyond \citet{gu2026prediction} for right-censored data, due to the complexity of the interval-censored likelihood, for which the relevant function classes may be unbounded and non-Donsker, and the additional challenges posed by the more delicate multi-source transfer learning procedure. 
We develop new mathematical arguments to address these difficulties. Finally, we demonstrate the performance of SPOT-IC through extensive simulation studies and an application to the Alzheimer's Disease Neuroimaging Initiative.

\section{Methods}\label{sec-meth}
\subsection{Model, data, and likelihood for the target study}
\label{subsec-target_model}
Suppose that the target study consists of $n$ independent subjects. For the $i$th subject $(i=1,\dots,n)$, let $T_i$ denote the failure time and $\X_i(\cdot)$ denote a $p$-dimensional vector of potentially time-dependent covariates. Under a class of semiparametric transformation models, the conditional cumulative hazard function of $T_i$ given $\X_i(\cdot)$ is
\begin{equation} \label{eq:trans_model}
    \Lambda_i(t|\X_i)=G\left[\int_0^t \exp\{\bb^{\T} \X_i(s)\}d\Lambda(s)\right],
\end{equation}
where $G(\cdot)$ is a prespecified, strictly increasing transformation function, $\bb$ is a $p$-dimensional vector of unknown regression coefficients, and $\Lambda(\cdot)$ is an unspecified nondecreasing function with $\Lambda(0) = 0$.

We consider a general mixed-case interval-censoring mechanism, in which both the number of examinations and the examination times may vary across subjects. Specifically, let $U_{i,1} < \dots < U_{i,D_i}$ denote the examination times for the $i$th subject, where $D_i$ is the total number of examinations for that subject. For convenience, define the augmented examination-time sequence $\tilde{\mbf{U}}_i = (U_{i,0}, U_{i,1}, \dots, U_{i,D_i}, U_{i,D_i+1})$, with $U_{i,0} = 0$ and $U_{i,D_i+1} = \infty$. The corresponding interval-specific event status vector is $\tilde{\mbf{\Delta}}_i = (\Delta_{i,0}, \Delta_{i,1}, \dots, \Delta_{i,D_i})$, where $\Delta_{i,d} = I(U_{i,d} < T_i \le U_{i,d+1})$ $(d = 0, \dots, D_i)$. Thus, the observed data from the target study are $\{\tilde{\mbf{U}}_i, \tilde{\mbf{\Delta}}_i, \X_i(t) : t\in [0,\tau], \,i=1,\dots,n\}$, where $\tau$ is the end of follow-up in the target study. 

We assume that the examination process, namely $(D_i, \tilde{\mbf{U}}_i)$, is independent of $T_i$ conditional on $\X_i(\cdot)$. Under this assumption, the observed time-status information for the $i$th subject can be fully represented by the shortest interval that brackets $T_i$, denoted by $(L_i,R_i]$, where
$L_i=\max\{U_{i,d}:U_{i,d}<T_i,\, d=0,\dots,D_i\}$ and $R_i = \min \{U_{i,d}:T_i\le U_{i,d},\, d=1,\dots,D_i+1\}$. The special cases $L_i=0$ and $R_i=\infty$ correspond to left-censoring and right-censoring, respectively.
Based on the simplified observed data $\{(L_i, R_i, \X_i): i=1,\dots,n\}$, the likelihood function concerning $(\bb, \Lambda)$ is
\begin{align*}
    L_n(\bb,\Lambda) = \prod_{i=1}^{n} &\Biggl\{
    \exp \left(-G\left[\int_0^{L_i} \exp\left\{\bb^{\T} \X_i(s)\right\} d\Lambda(s)\right]\right) \\
    & - \exp \left(-G\left[\int_0^{R_i} \exp\left\{\bb^{\T} \X_i(s)\right\} d\Lambda(s)\right]\right)
    \Biggr\}.
\end{align*}

% \begin{remark}
%     Complete observation of the full sequence of examination times is not necessary for inferential purposes, since only $L_i$ and $R_i$ are involved in the likelihood, and the remaining examination times make no contribution. Nevertheless, the theoretical analysis have to take into account the joint distribution for the entire sequence of examination times.
% \end{remark}

We specify $G(\cdot)$ via the frailty-induced log-Laplace transformation
\begin{equation*}
    G(x) = -\log \int_0^{\infty} \exp(-xz)f(z)dz,
\end{equation*}
where $f(z)$ is the density function of a nonnegative frailty variable $z$. Letting $z$ follow a gamma distribution with mean one and variance $r>0$ yields the logarithmic transformation family $G(x) = r^{-1}\log(1+rx)$, which includes the proportional odds model at $r=1$ and the Cox proportional hazards model by allowing $G(x) = x$ when $r=0$. Under the frailty-induced transformation model, the likelihood function can be written as
\begin{align*}
    L_n(\bb,\Lambda) = \prod_{i=1}^{n} \int_{z_i} \Biggl(&
    \exp \left[-\int_0^{L_i} z_i\exp\left\{\bb^{\T} \X_i(s)\right\} d\Lambda(s)\right] \\
    & - \exp \left[-\int_0^{R_i} z_i\exp\left\{\bb^{\T} \X_i(s)\right\} d\Lambda(s)\right]
    \Biggr)f(z_i)dz_i.
\end{align*}

% For each source study, SPOT-IC requires only an estimated survival function as auxiliary information. Suppose that there are $K$ source studies and let $\check{S}_k(t|\X_k)$ denote the predicted survival function for the $k$th source study, where $\X_k$ is the corresponding set of potentially time-dependent covariates, $k= 1,\dots,K$. These estimators are not subject to any specific modeling constraints and can be built using any statistical or machine learning method. Consequently, the covariates involved in the target study and all source studies are also allowed to differ. To facilitate information transfer from source studies, we assume that the target covariates $\X$ contain all the information needed to recover $\X_k$ $(k = 1,...,K)$ such that each $\X_k$ can be viewed as a function of $\X$. Therefore, we write the source estimators as $\check{S}(t|\X)$ hereafter, although it actually depends only on the covariates $\X_k$.

\subsection{SPOT-IC with single source}\label{subsec-single source}
We first consider the single-source setting.
Let $\check{S}(t|\X)$ denote the estimator of the conditional survival function given covariates $\X$, provided by the source study. This estimator can be constructed using any survival analysis techniques, including model-based, machine-learning, and AI methods, and can be based on any type of survival outcome in the source study, such as right-censored and interval-censored failure time data.
Moreover, the covariates required in $\check{S}$ need not align exactly with the target covariates $\X$; they can be any function of $\X$, such as a subset or transformation of $\X$. Thus, our approach provides substantial flexibility in the processing and sharing of source information compared to existing methods.

% For each source study, SPOT-IC requires only an estimated survival function as auxiliary information. Suppose that there are $K$ source studies and let $\check{S}_k(t|\X_k)$ denote the predicted survival function for the $k$th source study, where $\X_k$ is the corresponding set of potentially time-dependent covariates, $k= 1,\dots,K$. These estimators are not subject to any specific modeling constraints and can be built using any statistical or machine learning method. Consequently, the covariates involved in the target study and all source studies are also allowed to differ. To facilitate information transfer from source studies, we assume that the target covariates $\X$ contain all the information needed to recover $\X_k$ $(k = 1,...,K)$ such that each $\X_k$ can be viewed as a function of $\X$. Therefore, we write the source estimators as $\check{S}(t|\X)$ hereafter, although it actually depends only on the covariates $\X_k$.

% \subsubsection{Survival Probability Transfer}
% \label{subsubsec:spot}
The key idea of SPOT-IC is to incorporate a survival-probability similarity metric as a penalty in the objective function, thereby encouraging alignment between the target and source predictions. Specifically, let $S(t|\X; \bb,\Lambda) = \exp (-G[\int_0^t \exp\{\bb^{\T}\X(s)\} d\Lambda(s)])$ denote the survival function in the target study under model~\eqref{eq:trans_model}.
We quantify the similarity between the target survival function $S(t|\X)$ and the source estimator $\check S(t|\X)$ using the following empirical negative cross-entropy criterion:
\begin{equation*}
\psi_m({\bb}, \Lambda) = m^{-1} \sum_{i=1}^{m} 
\left[ \check{S}(\tilde{Y}_i | \tilde{\X}_i) \log S(\tilde{Y}_i | \tilde{\X}_i) 
+ \left\{ 1 - \check{S}(\tilde{Y}_i | \tilde{\X}_i) \right\} 
\log \left\{ 1 - S(\tilde{Y}_i | \tilde{\X}_i) \right\} \right],
\end{equation*}
where $m$ is a large positive integer, $\{\tilde{\X}_i: i= 1,...,m\}$ are independent copies of the target covariates $\X$, and $\{\tilde{Y}_i : i= 1,...,m\}$ are i.i.d. random variables whose support is the union of the supports of $(U_{i,1}, \dots, U_{i,D_i})$. In practice, $\tilde \X_i$ can be generated by resampling from the observed target covariates $\{\X_i:i=1,\dots,n\}$, while $\tilde{Y}_i$ can be drawn from a uniform or truncated exponential distribution.
% Let $\ell _n(\bb,\Lambda) = \log L_n(\bb,\Lambda)$ denote the log-likelihood function for the target study.  For notational simplicity, we suppress the subscript $k$ and write $\check S(t|\X)$ instead of $\check{S}_k(t|\X)$ in this setting. We 
It is easy to see that $\psi_m(\bb,\Lambda)$ attains its global maximum when $S(t|\X) = \check S(t|\X)$. This motivates using $\psi_m(\bb,\Lambda)$ as a penalty to shrink $S(t|\X)$ toward $\check S(t|\X)$. Accordingly, we estimate $(\bb,\Lambda)$ by solving 
\begin{equation}\label{eq:origin penalty}
(\hat{\bb}, \hat{\Lambda}) = \arg\max_{(\bb, \Lambda)} 
n^{-1} \log L_n(\bb, \Lambda) + \xi_n \psi_m(\bb, \Lambda),
\end{equation}
where $\xi_n\ge0$ is a tuning parameter controlling the degree of source information borrowing,
% The objective function \eqref{eq:origin penalty} will reduce to the target-only optimization problem when $\xi_n=0$, and will be dominated by the penalty term if $\xi_n = \infty$. 
and its optimal value can be selected using data-adaptive methods such as cross-validation.

A key advantage of the proposed cross-entropy-type penalty $\psi_m(\bb,\Lambda)$, compared to conventional  $L_q$-penalties ($q\in[0,2]$) on survival probabilities, is that it leads to much simpler computation as problem~\eqref{eq:origin penalty} can be solved by maximizing a weighted log-likelihood for interval-censored data. 
To illustrate this connection, we temporarily express $\psi_m(\bb,\Lambda)$ as
\begin{equation} \label{eq:psi_J}
\psi_m(\bb,\Lambda) = \lim_{J\to\infty} (mJ)^{-1} \sum_{i=1}^{m} \sum_{j=1}^{J}  
\left[ (1 - \delta_{ij}) \log S(\tilde{Y}_i | \tilde{\X}_i) 
+ \delta_{ij} \log \left\{ 1 - S(\tilde{Y}_i | \tilde{\X}_i) \right\} \right],
\end{equation}
where for each $i = 1,\dots,m$, $\{\delta_{ij}: j = 1,\dots,J\}$ denote a collection of i.i.d. Bernoulli random variables with mean $1-\check{S}(\tilde{Y}_i|\tilde{\X}_i)$.
The expression inside the limit in~\eqref{eq:psi_J} coincides with the weighted log-likelihood for the current status data $\{\tilde{Y}_{ij},\delta_{ij},\tilde{\X}_{ij}(t):t\in [0,\tau]\}$, with $\tilde{Y}_{ij} \equiv \tilde{Y}_i$ and $\tilde{\X}_{ij}\equiv\tilde{\X}_i$ $(j=1,\dots,J)$.
Therefore, the optimization problem in~\eqref{eq:origin penalty} falls within the maximum likelihood framework for interval-censored data and can be solved by extending the Poisson-data augmentation strategy of \citet{Zeng2016M} to the corresponding weighted likelihood setting.

As shown later in \eqref{eq:post-tilde-W} of the appendix, the optimization procedure depends on the artificial Bernoulli variables $\delta_{ij}$'s only through the average $J^{-1}\sum_{j=1}^J \delta_{ij}$, which converges to $1-\check{S}(\tilde{Y}_i|\tilde{\X}_i)$ as $J\rightarrow\infty$. Thus, the representation in~\eqref{eq:psi_J} is introduced only to motivate the methodology and is not used in the actual computation.

% \subsubsection{Optimization Procedure}
% \label{subsubsec:optim_procedure}
In the nonparametric maximum likelihood estimation, $\Lambda$ is treated as a step function with nonnegative jump sizes $\lambda_l$ at times $t_l$ ($l=1,\dots,L$), where $t_1<\dots<t_L$ denote the unique values of $\{L_i: L_i>0, i=1,\dots,n\}$, $\{R_i: R_i<\infty, i=1,\dots,n\}$, and $\{\tilde{Y_i}: i=1,\dots,m\}$. Then, the objective function in~\eqref{eq:origin penalty} can be written as 
\[
\begin{split}
  n^{-1} \sum_{i=1}^n \log  \Bigg[\int_{z_i} \exp\bigg(-\sum_{t_l\le L_i} z_i\lambda_le^{\bb^{\T} \X_{il}}\bigg) 
  \bigg\{ 1-\exp\bigg(-\sum_{L_i<t_l\le R_i} z_i\lambda_le^{\bb^{\T} \X_{il}}\bigg)\bigg\}^{I\{R_i<\infty\}}f(z_i)dz_i\Bigg] \\
 + \xi_n (mJ)^{-1} \sum_{i=1}^{m} \sum_{j=1}^{J} \log \Bigg[ \int_{\tilde z_i} 
\bigg\{ \exp\bigg( - \sum_{l : t_l \leq \tilde{Y}_i} \tilde{z}_i \lambda_l e^{{\bb}^{\T} \tilde{\X}_{il}} \bigg) \bigg\}^{1 - \delta_{ij}} \\
 \times \bigg\{ 1 - \exp\bigg( - \sum_{l : t_l \leq \tilde{Y}_i} \tilde{z}_i \lambda_l e^{{\bb}^{\T} \tilde{\X}_{il}} \bigg) \bigg\}^{\delta_{ij}} 
f(\tilde{z}_i) d\tilde{z}_i \Bigg],
\end{split}
\]
where $I(\cdot)$ denotes the indicator function, $z_i$ and $\tilde{z}_i$ are frailty variables arising from the frailty-induced transformation, $\X_{il}=\X_i(t_l)$, and $\tilde{\X}_{il}=\tilde{\X}_i(t_l)$. To achieve more tractable optimization, we introduce two sets of latent Poisson random variables corresponding to the primary log-likelihood term and the penalty term, respectively. Specifically, let $W_{il}$ $(i=1,\dots,n; \ l=1,\dots,L)$ and $\tilde{W}_{ijl}$ $(i=1,\dots,m; \ j=1,\dots,J;\ l=1,\dots,L)$ be independent Poisson variables with means $z_i\lambda_l\exp{(\bb^{\T}\X_{il})}$ and $\tilde{z}_i \lambda_l\exp{(\bb^{\T} \tilde{\X}_{il})}$, respectively. 
It is straightforward to verify that conditional on $z_i$, the likelihood contribution of the interval-censored data $(L_i,R_i,\X_i)$ is equivalent to the likelihood of the event
\[
\left\{\sum_{l:t_l \le L_i} W_{il} = 0\right\} \quad \bigcap \quad \left\{\sum_{l: L_i<t_l \le R_i}W_{il}>0\right\}^{I(R_i < \infty)}.
\]
%  $(A_i = 0, B_i > 0 : i = 1, \dots, n)$ takes the form 
% \begin{equation*}
%     \prod_{i=1}^n \int_{z_i} \left\{\prod_{t_l\le L_i} \pr(W_{il}=0|z_i)\right\}\left\{1-\pr(\sum_{L_i<t_l\le R_i} W_{il}=0|z_i)\right\}^{I\{R_i<\infty\}}f(z_i)dz_i,
% \end{equation*}
% which are equivalent to the likelihood for . 
Similarly, conditional on $\tilde{z}_i$, the likelihood contribution of the current status data $(\tilde{Y}_{ij},\delta_{ij},\tilde{\X}_{ij})$ is equivalent to the likelihood of the event $\sum_{l : t_l \leq \tilde{Y}_i} \tilde{W}_{ijl} > 0$ when $\delta_{ij} = 1$, and of the event $\sum_{l : t_l \leq \tilde{Y}_i} \tilde{W}_{ijl} = 0$ when $\delta_{ij} = 0$. 
Therefore, maximizing the objective function in~\eqref{eq:origin penalty} is equivalent to maximizing the weighted log-likelihood induced by these latent Poisson variables.
% \begin{equation*}
% \tilde{V}_{ij} = 
% \begin{cases}
% \sum_{l : t_l \leq \tilde{Y}_i} \tilde{W}_{ijl} > 0 & \text{if } \delta_{ij} = 1, \\
% \sum_{l : t_l \leq \tilde{Y}_i} \tilde{W}_{ijl} = 0 & \text{if } \delta_{ij} = 0,
% \end{cases}
% \quad \text{for } i = 1, \ldots, m \text{ and } j = 1, \ldots, J.
% \end{equation*}

We maximize the equivalent objective function via an EM algorithm, treating $z_i$, $\tilde{z}_i$, $W_{il}$ and $\tilde{W}_{ijl}$ as missing data. 
Technical details of the EM algorithm are provided in the appendix. 
We highlight two major advantages of our algorithm.
First, it depends on the artificial variables $\delta_{ij}$'s only through the posterior mean of $J^{-1}\sum_{j=1}^J \tilde{W}_{ijl}$, which eventually becomes a function of $1 - \check{S}(\tilde{Y}_i | \tilde{\X}_i)$. Consequently, neither the individual $\delta_{ij}$ nor the auxiliary quantity $J$ needs to be explicitly specified.
Second, the algorithm avoids inversion of large matrices as the high-dimensional parameters $\lambda_l$'s are updated explicitly in the M-step.

Let $(\hat\bb,\hat \Lambda)$ denote the final estimates of $(\bb,\Lambda)$. For the target population, the covariate-specific survival function $S(t|\X)$ is then estimated by $\hat S(t|\X)=S(t|\X;\hat \bb,\hat \Lambda)$.

\subsection{SPOT-IC with multiple sources}
\label{subsec-multi-source}
In some applications, multiple source studies are available, yet it is unclear which of them provide useful information for target predictions. It is therefore important to filter out or downweight noninformative sources to prevent negative transfer. The proposed multi-source SPOT-IC approach consists of two stages: a screening stage, in which we construct a collection of candidate estimators for the target survival function $S(t|\X)$ by applying the single-source SPOT-IC procedure introduced in $\mathsection$\ref{subsec-single source} to the target study and each source study separately; and an aggregation stage, in which we combine these candidate estimators via Q-aggregation to obtain a final estimator.

Specifically, suppose that there are $K$ source studies. We randomly split the target data into two subsets: a screening set indexed by $\mathcal{I}^{\mathrm{scr}}$ and an aggregation set indexed by $\mathcal{I}^{\mathrm{agg}}$. In the screening stage, we apply the single-source SPOT-IC procedure to the target samples in $\mathcal{I}^{\mathrm{scr}}$ together with the $k$th source study, yielding the survival function estimator $\hat S_k(t|\X)\ (k=1,\dots,K)$. We also construct a target-only estimator, denoted by $\hat S_0(t|\X)$, using only the target samples in $\mathcal{I}^{\mathrm{scr}}$. Let $\mathcal{S} = \{\hat S_0,\hat S_1,\ldots,\hat S_K\}$ denote the resulting collection of candidate estimators. 

In the aggregation stage, we consider convex combinations of the candidate estimators of the form $\hat S_{\bt}(t|\X) = \sum_{k=0}^K \theta_k \hat S_k(t|\X)$, where the aggregation weights $\bt = (\theta_0,\theta_1,\ldots,\theta_K)$ belong to the simplex $\mbf{\Theta} = \left\{(\theta_0,\theta_1,\ldots,\theta_K) \in \mathbb{R}^{K+1} : \theta_k \ge 0,\ \sum_{k=0}^K \theta_k = 1 \right\}$. Inspired by the Q-aggregation framework of \citet{Lecu2014optimal}, we define the following $Q$-function:
\begin{equation*}
\begin{aligned}
    Q({\bt};\mathcal{S}, \mathcal{I}^{\mathrm{agg}}) =
    &  -\frac{1}{2}\sum_{i \in \mathcal{I}^{\mathrm{agg}}} \log \left\{\hat S_{\bt}(L_i|\X_i)-\hat S_{\bt}(R_i|\X_i)\right\} \\
    &-\frac{1}{2}\sum_{k=0}^K \theta_k \sum_{i \in \mathcal{I}^{\mathrm{agg}}} \log \left\{\hat S_k(L_i|\X_i)-\hat S_k(R_i|\X_i)\right\}.
\end{aligned}
\end{equation*}
The optimal aggregation weights are obtained by minimizing the penalized $Q$-function:
\begin{equation} \label{eq-agg-weight}
\hat{\bt} = {\arg\min}_{\bt\in\mbf{\Theta}} \left\{ Q({\bt};\mathcal{S}, \mathcal{I}^{\mathrm{agg}}) + \xi_{\bt} \sum_{k=0}^K \theta_k \log \theta_k \right\},
\end{equation}
where $\xi_{\bt}>0$ is a tuning parameter, $|\mathcal{I}|$ denotes the cardinality of the set $\mathcal{I}$, and the entropy penalty stabilizes the optimization and prevents overly concentrated solutions. 
The final estimator for $S(t|\X)$ is $\hat{S}_{\hat \bt}(t|\X)$.
The complete multi-source SPOT-IC procedure is summarized in Algorithm~\ref{alg:SPOT-IC}.

\begin{remark}
The multi-source SPOT-IC procedure described above is based on a single sample split. In practice, to improve finite-sample stability and data efficiency, particularly when the target sample size is small, one may implement a cross-fitted version of the procedure by repeating the sample splitting multiple times and averaging the resulting estimators.
\end{remark}

\begin{algorithm}[ht]
\caption{Multi-Source SPOT-IC}
\label{alg:SPOT-IC}
\hspace*{\algorithmicindent} \textbf{Input:} Target data $\{(L_i, R_i, \X_i): i=1,\dots,n\}$ and $K$ source estimators $\{\check S_k\}_{k=1}^K$\\
\hspace*{\algorithmicindent} \textbf{Output:} Estimated survival function $\hat{S}_{\hat{\bt}}(t|\X)$
\begin{algorithmic}
\STATE \textbf{Step 1:} Randomly split the target samples into two sets indexed by $\mathcal{I}^{\mathrm{scr}}$ and $\mathcal{I}^{\mathrm{agg}}$.

\STATE \textbf{Step 2:} Obtain the target-only estimator $\hat S_0$ using the target samples in $\mathcal{I}^{\mathrm{scr}}$.

\STATE \textbf{Step 3:} 
\FOR{$k=1,\dots,K$}
\STATE Obtain the single-source estimator $\hat S_k$ by applying the single-source SPOT-IC procedure to the target samples in $\mathcal{I}^{\mathrm{scr}}$ together with the $k$th source study.
\ENDFOR

\STATE \textbf{Step 4:} For the candidate set $\mathcal{S} = \{\hat S_0,\hat S_1,\ldots,\hat S_K\}$, obtain the optimal aggregation weights by
% Randomly split fold $v$ into two equal subsets indexed by $\mathcal{I}_v^{\mathrm{tune}}$ and $\mathcal{I}_v^{\mathrm{agg}}$. Use subjects indexed by $\mathcal{I}_v^{\mathrm{tune}}$ to select $\xi_{\bt}$, and compute the objective function on subjects indexed by $\mathcal{I}_v^{\mathrm{agg}}$:
    \[
        \hat{\bt} = {\arg\min}_{\bt\in\mbf{\Theta}} \left\{ Q({\bt};\mathcal{S}, \mathcal{I}^{\mathrm{agg}}) + \xi_{\bt} \sum_{k=0}^K \theta_k \log \theta_k \right\},
    \]
    where the tuning parameter $\xi_{\bt}$ is selected via cross-validation.

\STATE \textbf{Step 5:} Output the aggregated estimator $\hat{S}_{\hat{\bt}} = \sum_{k=0}^K \hat{\theta}_k \hat{S}_k$.
\end{algorithmic}
\end{algorithm}

% Step $3$ adopts the Q-aggregation method \citep{Lecu2014optimal} with a Kullback-Leibler penalty. For each $v$, we consider the target distribution as a mixture of $\hat S_0^{(v)},\hat S_1^{(v)},\dots,\hat S_K^{(v)}$, with unknown mixing probabilities $\theta_0,\theta_1,\dots,\theta_K$. Estimating $\bt$ is a convex, parametric optimization problem. Since $\bt$ contains the optimal weights, the aggregated estimator should be closer to the truth than any individual estimator.

\section{Asymptotic theory}\label{sec-theory}
In this section, we establish the asymptotic theory under the general multi-source setting with $K$ source studies. The asymptotic results for the single-source setting are obtained as a byproduct, since the single-source SPOT-IC procedure is incorporated in the screening stage of multi-source SPOT-IC.  

For the target population, let $(\bb_0,\Lambda_0)$ denote the true value of $(\bb,\Lambda)$, and let $S_0(t|\X) = S(t|\X;\bb_0,\Lambda_0)$ denote the corresponding true survival function. For $k=1,\ldots,K$, let $S_k(t|\X)$ denote the true survival function for the $k$th source population. 
We also introduce the following empirical process notation. Let $\mbb P_n$ denote the empirical measure based on $n$ independent observations of interval-censored data $(\tilde{\mbf{U}}, \tilde{\mbf{\Delta}}, \X)$, and let $\mbb P$ denote the corresponding true probability measure. Likewise, for the penalty term, let $\tilde {\mbb P}_m$ denote the empirical measure based on $m$ independent observations of $(\tilde Y,\tilde \X)$, and let $\tilde {\mbb P}$ denote the corresponding true probability measure. Then the objective function in \eqref{eq:origin penalty} can be written as $\mbb P_n \ell(\bb,\Lambda)+\xi_n\tilde{\mbb P}_m \psi(\bb,\Lambda)$, where
$\ell(\bb,\Lambda)$ and $\psi(\bb,\Lambda)$ denote the log-likelihood function and the penalty from a single observation, respectively.

We establish the asymptotic theory under the following conditions, where we consider a generic subject and omit the subscript $i$ from all random quantities.

\begin{condition}\label{con1:true-param}
The true value ${\bb}_0$ lies in the interior of a known compact set $\mathcal B$ in $\mathbb R^p$, and the true value $\Lambda_0(\cdot)$ is continuously differentiable with positive derivatives in $[\zeta, \tau]$, where $[\zeta, \tau]$ is the union of the supports of $(U_1, \dots, U_D )$.
\end{condition}

\begin{condition}\label{con2:differen-X}
With probability one, the covariate process $\X(t)$ is continuously differentiable in $[\zeta, \tau]$. Moreover, if there exist a constant vector $\mbf{a}_1$ and a deterministic function $a_2(t)$ such that for any $t \in [\zeta,\tau]$, $\mbf{a}_1^{\T} \X(t) + a_2(t) = 0$ with  probability 1, then $\mbf{a}_1 = \mbf 0$ and $a_2(t) = 0$ for $t\in[\zeta, \tau]$.
\end{condition}

\begin{condition}\label{con3:exam-times}
The number of examination times $D$ is positive with $E(D) < \infty$. %The conditional probability $\Pr(U_D = \tau |D ,X)$ is greater than some positive constant c. 
In addition, $\pr\{\min_{0\le d\le D}(U _{d+1}-U_{d})\ge \gamma | D, \X \} = 1$ for some positive constant $\gamma$.
Finally, the conditional densities of $(U_d,U_{d+1})$ given $\X$ and $D$, denoted by $g_d(u,v|\X,D)$ ($d=1,\dots,D$), have continuous partial derivatives with respect to $u$ and $v$ when $v - u\ge \gamma$, and are continuously differentiable functionals with respect to $\X$. 
\end{condition} 

\begin{condition} \label{con4:G-func}
The transformation function $G(\cdot)$ is twice continuously differentiable on $[0, \infty)$ with $ G(0) = 0$, $G'(x) > 0$, and $G(\infty) = \infty$. 
\end{condition}

\begin{condition} \label{con5:converge-rate-S-check}
For $k=1,\dots,K$, the source estimator $\check S_k(t|\X)$ is well-defined for any $\X$ from the target population and any $t \in [0,\tau]$. Let $N_k$ denote the sample size of the $k$th source study. There exists a sequence $q_k \to 0$ as $N_k \to \infty$ such that
$$E\|\check
S_k(t|\X)- S_k(t|\X)\|^2
_{L_2} \le O_p(q_k),$$
where $\|\cdot\|_{L_2}$ denotes the $L_2$-norm over $[\zeta,\tau]$, and the expectation is taken with respect to the distribution of $\X$ in the target population. 
\end{condition}

\begin{condition} \label{con6:tilde-Y-X}
The conditional density function of $\tilde Y$ given $\tilde \X$ is uniformly bounded almost everywhere in its support $[\zeta,\tau]$. % and its support $[\tilde \zeta,\tau]$ satisfies $ [\zeta,\tau]\subsetneq [\tilde \zeta,\tau]$.
In addition, with probability one, %$\tilde Y$ is bounded away from 0, 
% $w(\tilde Y,\tilde \X)$ is bounded away from both $0$ and $\infty$, and 
$\check S_k(\tilde Y|\tilde \X)$ $(k=1,\dots,K)$ is bounded away from both $0$ and $1$.
\end{condition}

\begin{condition} \label{con7:identi-obj}
For any $\xi \in (0,\infty]$, with $\xi^{-1}$ defined as $0$ when $\xi=\infty$, the function $\xi^{-1} \mbb P \ell({\bb},\Lambda) + \tilde {\mbb P} \psi({\bb},\Lambda)$ admits a unique maximizer $({\bb}^*,\Lambda^*)$ such that ${\bb}^*$ lies in the interior of $\mathcal B$ and $\Lambda^*$ is a non-decreasing function with $\Lambda^*(0) = 0$ and $\Lambda^*(\tau) < \infty$. Here, the uniqueness of $\Lambda^*$ holds on $[\zeta,\tau]$. In addition, $\Lambda^*(t)$ is continuously differentiable with positive derivatives in $[\zeta,\tau]$.
\end{condition}

\begin{remark}
    % Conditions~\ref{con1:true-param} and \ref{con2:differen-X} are standard regularity conditions in regression analysis of survival outcomes. Condition~\ref{con3:exam-times} pertains to the joint distribution of the examination times. It requires that the gap between any two consecutive examination times be no smaller than a positive constant $\gamma$, thereby excluding exact observations, which require different arguments. The smoothness assumption on the joint density of the examination times is used to prove the Donsker property of certain function classes. Compared with Condition 4 of \citet{Zeng2016M}, we do not require the largest examination time to be exactly $\tau$ with positive probability.
    Conditions~\ref{con1:true-param}--\ref{con3:exam-times} are standard regularity conditions in semiparametric regression analysis of mixed-case interval-censored data \citep{Zeng2016M,zeng2017multivariate}. Condition~\ref{con4:G-func} is satisfied by the logarithmic transformation models introduced in $\mathsection$\ref{subsec-target_model}. Condition~\ref{con5:converge-rate-S-check} specifies the convergence rate of each source estimator $\check S_k(t|\X)$, which generally depends on the data type and estimation procedure used in the source study. For example, for right-censored source data, conventional semiparametric regression methods, such as Cox regression and transformation models, typically achieve the parametric rate $q_k = N_k^{-1}$, while modern machine learning methods generally converge at slower rates. For interval-censored source data, the convergence rate of $\check S_k$ is typically slower than $N_k^{-1}$ and can attain the rate $q_k = N_k^{-2/3}$ under suitable smoothness conditions \citep{huang1996efficient,huang1997sieve,Zeng2016M}. Condition~\ref{con6:tilde-Y-X} holds under a variety of data-generating mechanisms, such as those described in $\mathsection$\ref{subsec-single source}. Condition~\ref{con7:identi-obj} is an identifiability condition for the limiting objective function. 
    % We only require $\Lambda^*$ be uniquely defined on $[\zeta,\tau]$, which is consistent with the fact that the nonparametric component in transformation models for interval-censored data is identifiable only over the support of the examination times \citep{Zeng2016M}.
\end{remark}

We first establish the convergence rate for each candidate estimator $\hat S_k(t|\X)$ ($k=1,\dots,K$) obtained in the screening stage of multi-source SPOT-IC. We assume that in the sample-splitting step, the screening set $\mathcal I^{\mathrm{scr}}$ satisfies $|\mathcal{I}^{\mathrm{scr}}|/n\to c_0\in (0,1)$ as $n\to\infty$. This unifies the asymptotic analysis for single-source SPOT-IC based on either the entire target dataset or only the subset indexed by $\mathcal{I}^{\mathrm{scr}}$. In addition, with a slight abuse of notation, we use $\xi_n$ to denote a generic tuning parameter sequence, whose value and limit may vary across $k=1,\ldots,K$. 

\begin{theorem} \label{thm-single-source}
Suppose that $\xi_n \to \xi$ as $n \to \infty$. Under Conditions~\ref{con1:true-param}--\ref{con7:identi-obj}, for $k=1,\dots,K$,
$$
E\|\hat S_k(t| \X) - S_0(t| \X)\|_{L_2}^2
\le
\begin{cases}
O_p\!\left\{
n^{-2/3}
+ \xi_n^{4/3}m^{-2/3}
+ (\xi_n - \xi)^2
+ \xi (\eta_k + q_k)\right\}, & \text{if } \xi < \infty, \\[1em]
O_p\!\left(\xi_n^{-4/3}n^{-2/3}+
\xi_n^{-2}
+ m^{-2/3}
+ \eta_k
+ q_k
\right), & \text{if } \xi = \infty,
\end{cases}
$$
where $\eta_k = E\|S_k(t|\X)-S_0(t|\X)\|_{L_2}^2$.
\end{theorem}

% We further define the distance between $S^s(t|\X)$ and $S_0(t|\X)$ by
% $$\eta_k = E\|S^s_k(t|\X)-S_0(t|\X)\|_{L_2}^2.$$ 
% In addition, let $\eta = \min_{k=1,\dots,K} \eta_k$.

% For the single-source case, we write $S_k(t\mid\X)$ and $\check S_k(t\mid\X)$ as $S^s(t\mid\X)$ and $\check S(t\mid\X)$, respectively. Clearly, $\eta = E\|S^s(t|\X)-S_0(t|\X)\|_{L_2}^2$. 
% Theorem 1 specifies the convergence rates of the proposed estimator $\hat S(t|\X)$ under two regimes of $\xi_n$ in the single-source setting.

% \begin{remark}
% Since $m$ can be taken arbitrarily large in our setting, terms involving $m^{-2/3}$ in the above rate can become negligible. The convergence rate of $\hat S(t|\X)$ is therefore jointly determined by the sample size of the target study, the bias introduced by the source estimator, and the tuning parameter $\xi_n$, where $\xi_n$ determines the extent to which the source information is borrowed. In particular, if $\xi_n$ diverges sufficiently fast, the convergence rate of $\hat S(t|\X)$ will then be determined by the bias term $(q_N+\eta)$.
% \end{remark}

\begin{remark}
    The proof of Theorem~\ref{thm-single-source} requires establishing upper bounds for $E\|\hat S_k(t|\X)-S^*(t|\X)\|_{L_2}^2$ and $E\|S^*(t|\X)-S_0(t|\X)\|_{L_2}^2$, where $S^*(t|\X) = S(t|\X;{\bb}^*,\Lambda^*)$. Although this high-level error decomposition has been used in \citet{gu2026prediction} for right-censored data, establishing these bounds under interval censoring requires fundamentally different arguments. In particular, our proof introduces two major technical innovations. First, the interval-censored log-likelihood and the penalty term can be unbounded. To achieve the Donsker property for relevant function classes, we consider convex combinations of the likelihood contributions under arbitrary $(\bb, \Lambda)$ and $(\bb^*, \Lambda^*)$. The consistency and convergence rate are then established through new arguments based on Fatou's lemma and Hellinger distance, respectively. Second, when $\xi=\infty$, the limiting objective function reduces to the penalty term alone. Standard $M$-estimation theory is inapplicable in this scenario because the maximizer $S^*(t|\X)$ may not coincide with $\check S_k(t|\X)$, the true survival function in the associated pseudo population. To overcome this difficulty, we carefully analyse the fluctuation of the limiting objective function and derive an upper bound in terms of $E\|\check S_k(t|\X)-S^*(t|\X)\|_{L_2}^2$, which motivates a novel use of $M$-estimation theory.
\end{remark}

We further optimize the convergence rate in Theorem~\ref{thm-single-source} with respect to $\xi_n$ while letting $m \to \infty$. The following theorem establishes the optimal convergence rate for each candidate estimator $\hat S_k(t| \X)$, together with a theoretically justified choice of $\xi_n$. 
% Let $\hat S^{tar}(t|\X)$ denote the survival estimator from the target-only method. Theorem 2 further provides a theoretically justified choice of $\xi_n$, under which $\hat S(t|\X)$ attains a rate no slower than that of the target-only estimator.
\begin{theorem} \label{thm-opt-rate}
Under Conditions \ref{con1:true-param}--\ref{con7:identi-obj}, the optimal convergence rate of $\hat{S}_k(t|\X)$ is
$$
E\|\hat{S}_k(t|\X) - S_0(t|\X)\|_{L_2}^2 \le O_p \left\{ n^{-2/3} \wedge (\eta_k + q_k) \right\}, \quad\text{ for } k=1,\dots,K,
$$
where $x \wedge y = \min(x,y)$. For suitable positive constants $\nu$ and $c$ such that $n^{-2(1+\nu)/3}+n^{-\nu} \le O(q_k)$, the optimal value of $\xi_n$ can be chosen as 
$$\xi_n^{\mathrm{opt}} = \Big\{\tilde{\mbb P}_m \|\check S_k(t|\X)-\hat S_0(t|\X)\|_{L_2}^2\Big\}^{-3\nu/4}I\Big\{\tilde{\mbb P}_m \|\check S_k(t|\X)-\hat S_0(t|\X)\|_{L_2}^2+q_k \le c n^{-2/3}\Big\},$$
where $\hat S_0(t|\X)$ is the target-only estimator obtained in Step 2 of Algorithm~\ref{alg:SPOT-IC}.
\end{theorem}

\begin{remark}
    An important implication of Theorem~\ref{thm-opt-rate} is that, in the single-source setting, SPOT-IC can achieve a faster convergence rate than the target-only benchmark $O_p(n^{-2/3})$ established in \citet{Zeng2016M}, provided that the target and source survival functions are sufficiently similar within the target support and the source estimator converges sufficiently fast so that $\eta_k + q_k = o(n^{-2/3})$. In practice, $\xi_n^{\mathrm{opt}}$ in Theorem~\ref{thm-opt-rate} may provide guidance for constructing candidate values of $\xi_n$ in cross-validation.
\end{remark}

% \begin{remark}
%    If the source estimator satisfies $q_N+\eta = o(n^{-2/3})$, SPOT-IC can achieve a faster rate than the target-only benchmark $O_p(n^{-1/3})$. In our numerical studies, once $\xi_n$ is sufficiently large, the prediction performance is relatively insensitive to further increases in $\xi_n$. $\xi_n$ selected via cross-validation also leads to satisfying results. 
% \end{remark}

Finally, we establish the convergence rate for the aggregated estimator $\hat S_{\hat{\bt}}(t|\X)$.
\begin{theorem} \label{thm-multi-rate}
Suppose that the tuning parameter $\xi_{\bt}$ in \eqref{eq-agg-weight} satisfies $\xi_{\bt} \ge C_0$, where $C_0$ is a positive constant specified in \eqref{eq:thm3-C(b)} of the Supplementary Material. Under Conditions~\ref{con1:true-param}--\ref{con7:identi-obj}, 
$$E\|\hat S_{\hat {\bt}}(t|\X) -S_0(t|\X)\|_{L_2}^2 \le O_p\left[n^{-2/3} \wedge \left\{\min_{1\le k \le K}(\eta_k+q_k)+n^{-1}\right\}\right].$$
\end{theorem}
% \begin{theorem} \label{thm-multi-rate}
% Assume that the tuning parameter $\xi_{\bt}$ in \eqref{eq-agg-weight} satisfies $\xi_{\bt} \ge C(b)$, where $b$ is a positive constant such that 
% \[
% \min_{\substack{
% 0 \le k \le K\\
% 0 \le d \le D
% }}\left|\hat S_k(U_d|\X)-\hat S_k(U_{d+1}|\X)\right| \ge b
% \]
% almost surely for sufficiently large $n$, and $C(b)$ is a positive constant depending on $b$, as defined in \eqref{eq:thm3-C(b)} of the supplementary materials. Then, under Conditions~\ref{con1:true-param}--\ref{con7:identi-obj}, 
% % it holds that
% % $$E\|\hat S_{\hat {\bt}}(t|\X) -S_0(t|\X)\|_{L_2}^2 \le \min_{j=0,1,\dots,K} E \|\hat S_j(t|\X)-S_0(t|\X)\|_{L_2}^2 +O_p(n^{-1}).$$
% % Consequently, the convergence rate of $\hat S_{\hat {\bt}}(t|\X)$ satisfies 
% $$E\|\hat S_{\hat {\bt}}(t|\X) -S_0(t|\X)\|_{L_2}^2 \le O_p\left[n^{-2/3} \wedge \left\{\min_{1\le k \le K}(\eta_k+q_k)+n^{-1}\right\}\right].$$
% \end{theorem}

\begin{remark}
    %The consistency of the candidate estimators $\hat{S}_k$ ($k=1,\dots,K$) established in Theorem~\ref{thm-opt-rate}, together with Conditions~\ref{con1:true-param} and \ref{con3:exam-times}, guarantees the existence of the constant $b$ in Theorem~\ref{thm-multi-rate}. 
    Theorem~\ref{thm-multi-rate} shows that the aggregated estimator $\hat S_{\hat{\bt}}$ achieves, up to an additive term of order $O_p(n^{-1})$, the convergence rate of the best candidate estimator. Consequently, $\hat S_{\hat{\bt}}$ attains a faster convergence rate than the target-only estimator whenever at least one candidate estimator does. This result parallels the optimal oracle inequalities established in the Q-aggregation literature \citep{dai2012deviation,Lecu2014optimal}. However, existing Q-aggregation theory does not directly apply because our penalized $Q$-function differs from the classical formulation and, more fundamentally, oracle inequalities for aggregation risk do not generally imply estimation error bounds for the aggregated estimator. We bridge this gap by developing novel arguments based on the equivalence of norms on the underlying Banach space.
\end{remark}

\section{Simulation studies}\label{sec-simu}
% In this section, we evaluate the prediction performance of SPOT-IC through comprehensive comparisons with existing methods in both the single-source and multi-source settings.
% % via various simulation studies. The first study considers a single source, while the second allows multiple sources, some closely aligned with the target and others more divergent.

\subsection{Single-source setting} 
\label{subsec-single-sim}
To evaluate the prediction performance of SPOT-IC, we conduct comprehensive comparisons with existing methods in both single-source and multi-source settings. We begin with the single-source setting. For the target study, we generate two independent covariates, $X_1 \sim \mathrm{Unif}(0,1)$ and $X_2 \sim \mathrm{Ber}(0.5)$, and generate the failure time from a Cox model with cumulative hazard
\[
    \Lambda(t|X_1,X_2) = \Lambda_0(t)\exp(\beta_1X_1+\beta_2X_2),
\]
where $\Lambda_0(t) = \log(1+0.5\sqrt t)$ and $(\beta_1,\beta_2) = (0.5,-0.7)$.
We set the target sample size to $n= 100$ and the study end time to $\tau = 2$.
For each subject, we randomly generate four examination times sequentially as follows: $U_1\sim\mathrm{Exp}(4)2\tau/3$ and $U_d = U_{d-1} + 0.1 + \mathrm{Exp}(4)2\tau/3$ $(d=2,3,4)$; any examination time exceeding $\tau$ is truncated at $\tau$. There are about 65\% right-censored observations and 20\% left-censored ones.

We consider a source study with 1,000 subjects and study duration $\tau_s = 5$. Its covariates are generated the same way as in the target study. We consider the following five scenarios, each having a distinct source model:
\begin{itemize}
\item[] Scenario 1: The source model is identical to the target model.

\item[] Scenario 2: The source model has a different cumulative baseline hazard, $\Lambda_0(t) = \log(1+0.6\sqrt t)$.

\item[] Scenario 3: The source model has different cumulative baseline hazard and regression coefficients, with $\Lambda_0(t) = \log(1+0.6\sqrt t)$ and $(\beta_1,\beta_2) = (0.7,-1)$. 

\item[] Scenario 4: The source model follows a proportional odds model with cumulative hazard
$$\Lambda(t|X_1,X_2)  = \log\left\{1+0.6\sqrt t\exp(0.7X_1-X_2)\right\}.$$

\item[] Scenario 5: The source model follows an accelerated failure time model $$\log T = 2 \left\{-0.7X_1+X_2 + W + N(0,0.5^2)\right\},$$ 
where $W$ is a random variable with survival function $S_W(w)=(1 + 0.6e^w)^{-1}$.
\end{itemize}
We generate eight examination times for each subject in the source study, with $U_1 \sim \mathrm{Exp}(4)\tau_s/3$ and $U_d =  U_{d-1} + 0.1 + \mathrm{Exp}(4)\tau_s/3$ $(d=2,\dots,8)$. Again, all examination times exceeding $\tau_s$ are truncated at $\tau_s$. On average, about 50--55\% of the observations are right-censored and 20--25\% are left-censored.
% In Scenario 1, approximately $50\%$ of the observations are right-censored, and $20\%$ are left-censored. In Scenarios 2--5, the rates of right- and left-censoring are approximately $55\%$ and $25\%$, respectively.

To obtain a source estimator, we fit a logarithmic transformation model with transformation function $G(x) = r^{-1}\log(1 + rx)$, where $r$ is determined through a grid search using the Akaike information criterion (\AIC).

For comparison, we consider four competing methods:
the target-only method, a combined method that pools target and source data, the Trans-Cox-MIC method developed by \citet{xie2024transfer}, and a lasso-based transfer learning method (Trans-Cox-Lasso). As discussed in \citet{xie2024transfer}, Trans-Cox-Lasso replaces the $L_0$-penalty in Trans-Cox-MIC with the lasso penalty and can be viewed as an extension of the Trans-Cox method of \citet{lizy2023accommodating} from right-censored data to interval-censored data. The target-only, combined, and our SPOT-IC methods adopt logarithmic transformation models, with the optimal transformation parameter selected via \AIC. The tuning parameter $\xi_n$ in SPOT-IC is selected via 5-fold cross-validation.

A validation dataset comprising $n_v = 10,000$ uncensored samples is generated from the true target model to evaluate the prediction performance. Each scenario is replicated $1,000$ times. The prediction performance is evaluated using the following four metrics:
\begin{itemize}
\item[] $L_2$D: The $L_2$-distance between $\hat{S}(t|X)$ and $S_0(t|X)$ over $[0,\tau]$.

\item[] SupAE: The supremum absolute error between $\hat{S}(t|X)$ and $S_0(t|X)$ over $[0,\tau]$.

\item[] IBS: The integrated Brier score \citep{graf1999assessment} over $[0,\tau]$, defined as 
$$
\mathrm{IBS}(\hat S)=\frac{1}{n_v}\sum_{i=1}^{n_v} \frac{1}{\tau}\int_0^\tau \left\{ I(T_i>t)-\hat S(t | \X_i)\right\}^2\,dt.
$$

\item[] C-index: The concordance index \citep{uno2011cindex}.
\end{itemize}
\begin{remark}
    A larger C-index indicates better performance, whereas smaller values of all other metrics indicate better performance. Since the validation data in this section are uncensored, the indicator $I(T_i > t)$ in the IBS metric is straightforward. In applications with interval-censored validation data, such as the one considered in $\mathsection$\ref{sec-app}, $I(T_i>t)=1$ if $t\le L_i$ and $I(T_i>t)=0$ if $t>R_i$. For $t\in(L_i,R_i]$, the indicator can be replaced by its estimated conditional expectation given $T_i\in(L_i,R_i]$, computed using an appropriate survival function estimator such as the Turnbull estimator \citep{turnbull1976empirical}. 
% $
% \{\hat S(t | \X_i)^{TB}-\hat S^{TB}(R_i | \X_i)\}/
% \{\hat S^{TB}(L_i | \X_i)-\hat S^{TB}(R_i | \X_i)\}$, where $\hat S^{TB}(t|\X)$ denotes the Turnbull estimator. 
\end{remark}

Table \ref{table:simu} summarizes the performance of all methods across the five scenarios. 
\begin{table}
	\tbl{Prediction results in the single-source simulation study}
{	\begin{tabular}{llccccc}
		Scenario & Metric & SPOT-IC & Target-only & Combined & Trans-Cox-Lasso & Trans-Cox-MIC \\
		1 & $L_2$D   & 0.048 (0.029) & 0.106 (0.037) & 0.031 (0.010) & 0.065 (0.024) & 0.066 (0.021) \\
		  & SupAE    & 0.053 (0.019) & 0.112 (0.023) & 0.049 (0.011) & 0.074 (0.008) & 0.076 (0.011) \\
		  & IBS      & 0.195 (0.002) & 0.199 (0.004) & 0.194 (0.002) & 0.196 (0.003) & 0.196 (0.002) \\
		  & C-index  & 0.604 (0.003) & 0.600 (0.009) & 0.604 (0.003) & 0.603 (0.004) & 0.604 (0.003) \\
		&   &  &  &  &  &  \\
		2 & $L_2$D   & 0.050 (0.023) & 0.106 (0.037) & 0.057 (0.015) & 0.081 (0.030) & 0.066 (0.022) \\
		  & SupAE    & 0.052 (0.020) & 0.112 (0.023) & 0.064 (0.012) & 0.074 (0.012) & 0.076 (0.011) \\
          & IBS      & 0.195 (0.002) & 0.199 (0.004) & 0.196 (0.002) & 0.197 (0.003) & 0.196 (0.002) \\
          & C-index  & 0.604 (0.003) & 0.600 (0.009) & 0.604 (0.003) & 0.603 (0.005) & 0.604 (0.003) \\
		&   &  &  &  &  &  \\
		3 & $L_2$D   & 0.067 (0.035) & 0.106 (0.037) & 0.082 (0.017) & 0.093 (0.028) & 0.084 (0.020) \\
          & SupAE    & 0.057 (0.022) & 0.112 (0.023) & 0.065 (0.011) & 0.074 (0.013) & 0.075 (0.010) \\
          & IBS      & 0.196 (0.003) & 0.199 (0.004) & 0.197 (0.002) & 0.198 (0.003) & 0.197 (0.003) \\
          & C-index  & 0.604 (0.003) & 0.600 (0.009) & 0.604 (0.003) & 0.603 (0.005) & 0.604 (0.003) \\
		&   &  &  &  &  &  \\
		4 & $L_2$D   & 0.049 (0.022) & 0.106 (0.037) & 0.056 (0.017) & 0.077 (0.027) & 0.066 (0.020) \\
        & SupAE    & 0.051 (0.017) & 0.112 (0.023) & 0.061 (0.012) & 0.071 (0.009) & 0.076 (0.011) \\
        & IBS      & 0.195 (0.002) & 0.199 (0.004) & 0.195 (0.002) & 0.197 (0.003) & 0.196 (0.002) \\
        & C-index  & 0.604 (0.003) & 0.600 (0.009) & 0.604 (0.003) & 0.603 (0.004) & 0.604 (0.003) \\
		&   &  &  &  &  &  \\
		5 & $L_2$D   & 0.052 (0.026) & 0.106 (0.037) & 0.061 (0.016) & 0.079 (0.027) & 0.066 (0.021) \\
        & SupAE    & 0.054 (0.020) & 0.112 (0.023) & 0.067 (0.013) & 0.072 (0.010) & 0.075 (0.011) \\
        & IBS      & 0.195 (0.002) & 0.199 (0.004) & 0.196 (0.002) & 0.197 (0.003) & 0.196 (0.002) \\
        & C-index  & 0.604 (0.003) & 0.600 (0.009) & 0.604 (0.003) & 0.603 (0.004) & 0.604 (0.003) \\
	\end{tabular}}
	\label{table:simu}
    \begin{tabnote}
        Median (median absolute deviation) over 1000 replicates are reported.
    \end{tabnote}
\end{table}
In Scenario 1, as expected, the combined method performs best, as the source and target studies share the same survival distribution and individual-level source data are fully utilized. Our SPOT-IC method ranks second and yields smaller median prediction errors than the other three methods. In Scenarios 2--5, the source distribution progressively departs from the target and may introduce misleading information. SPOT-IC consistently outperforms all other methods, and its advantage is particularly evident in the $L_2$D and SupAE metrics, both of which directly quantify the discrepancy between the estimated and true survival functions. 
% In Scenarios 4--5 where the source data no longer follow the Cox model, Trans-Cox-Lasso and Trans-Cox-MIC suffer from source model misspecification and thus perform badly. However, SPOT-IC does not impose such a structural assumption and thus remains robust.
These results demonstrate that SPOT-IC transfers source information more effectively and is more robust to negative transfer than existing methods.

\subsection{Multi-source setting}
\label{subsec-multi-sim}
In this subsection, we consider a multi-source setting with $K = 10$ source studies. Among them, $K_1$ sources are informative in the sense that their covariate-specific survival functions are similar to that of the target population, whereas the remaining $K-K_1$ sources are noninformative and differ substantially from the target. The informativeness of each source study is assumed to be unknown.

We set the target sample size to 150 and the sample size of each source study to 500. The target data are generated in the same way as in $\mathsection$\ref{subsec-single-sim}. For $k=1,\dots,K$, the $k$th source dataset is generated from a Cox model with cumulative hazard $\Lambda^{(k)}(t | \X) = \Lambda_0^{(k)}(t) \exp\{\bb^{(k)^{\T}}\X\}$, where the source covariates $\X = (X_1,X_2)^{\T}$ are generated as in $\mathsection$\ref{subsec-single-sim}. For the $K_1$ informative sources, we set 
$$\Lambda_0^{(k)}(t) = \log\left\{1+\left(0.5+0.05  B^{(k)}_0\right)\sqrt{t}\right\}, \quad \bb^{(k)} = (0.5, -0.7)^{\T} + 0.05 {\mbf{B}}^{(k)},$$ 
where $\mbf{B}^{(k)} = (B_1^{(k)},B_2^{(k)})$, and $B_0^{(k)}$, $B_1^{(k)}$, $B_2^{(k)}$ are independent Rademacher random variables. For the remaining $K-K_1$ noninformative sources, we set 
$$\Lambda_0^{(k)}(t)=\log \left\{ 1+\left(0.5+0.25B_0^{(k)} \right) \sqrt{t}\right\}, \quad \bb^{(k)} = (0.5, -0.7)^{\T} + 0.4\mbf{B}^{(k)},$$ 
where $\mbf{B}^{(k)} = (B_1^{(k)},B_2^{(k)})$, $\pr(B^{(k)}_0=-1)=0.7$, $\pr(B^{(k)}_0=1)=0.3$, and $B_1^{(k)}$ and $B_2^{(k)}$ are i.i.d with $\pr(B^{(k)}_j=-1)=0.7$ and $\pr(B^{(k)}_j=0)=0.3$ $(j=1,2)$. All random variables in $\{B^{(k)}_0, B^{(k)}_1, B^{(k)}_2: k=1,\dots, K\}$ are mutually independent.

We compare SPOT-IC with five competing methods: the target-only method; Oracle-SPOT-IC, which performs the SPOT-IC procedure using only the $K_1$ informative sources; Naive-SPOT-IC, which differs from SPOT-IC by assigning equal weights to all $K+1$ candidate estimators instead of using Q-aggregation; and TSDE and MAE, two multi-source transfer learning methods proposed by \citet{lou2026two}. 
% Oracle-SPOT-IC is obtained using the same procedure as SPOT-IC, but only similar sources ($k=1,\dots,K_1$) are used to construct candidate estimators. It reduces to the Target-only estimator when $K_1 = 0$. Naive-SPOT-IC combines all $K+1$ estimators with equal aggregation weights. TSDE and MAE proposed by \citet{lou2026two} are two source-detection-based transfer learning methods that extent \citet{xie2024transfer}. 

\begin{figure}[ht]
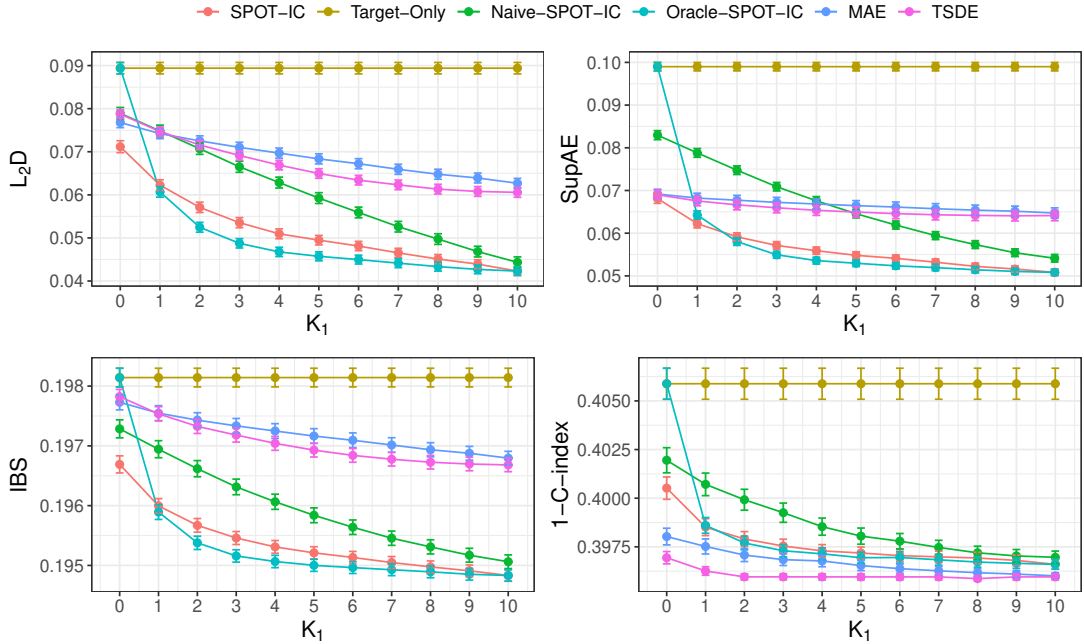

% The arguments in the next line are {height}{optional width}{used only by OUP for typesetting} for figure empty box eg 
%\figurebox{20pc}{25pc}{}
%if actual size of graphics need plese see below command
%\figurebox{}{}{}[fig1]
%need to reducing the figure size use below command
%\figuresize{.8}%
\figuresize{.6}
\figurebox{20pc}{25pc}{}[simu]
\caption{Prediction results in the multi-source simulation study under varying numbers of informative sources. Points and error bars represent the mean and mean $\pm$ standard deviation across 500 replicates, respectively.}
\label{fig:multiQ}
\end{figure}

We evaluate the prediction performance of each method using the same four metrics as in $\mathsection$\ref{subsec-single-sim}. Figure~\ref{fig:multiQ} presents the prediction results over 500 replicates.

As expected, for all methods except the target-only method, prediction error decreases as $K_1$, the number of informative sources, increases. When $K_1\ge 1$, Oracle-SPOT-IC consistently achieves the best performance because it has access to the true informativeness of each source study and excludes all noninformative sources. SPOT-IC closely tracks the performance of Oracle-SPOT-IC and consistently outperforms the remaining methods. This suggests that the proposed aggregation procedure in SPOT-IC can effectively identify and leverage informative source estimators while downweighting noninformative ones. 
An interesting phenomenon arises when no informative source is available (i.e., $K_1=0$). In this case, the target-only method and Oracle-SPOT-IC (which reduces to the target-only method) perform worse than the transfer learning methods. A possible explanation is that transfer learning methods based on model averaging or pooled source data analysis can partially offset biases from individual source studies when these biases operate in different directions, resulting in a final estimator with smaller overall bias than the target-only estimator. In particular, SPOT-IC performs best because its aggregation weights are chosen in a data-adaptive manner.
% We also observe that when $K_1=0$, SPOT-IC and naive-SPOT-IC yield smaller prediction errors than the target-only and Oracle-SPOT-IC where the latter two do not use auxiliary information. This suggests that despite the large bias in individual sources, SPOT-IC can offset opposing biases through weighted aggregation to some extent and some auxiliary information can still be borrowed.

\section{Application}\label{sec-app}
The Alzheimer’s Disease Neuroimaging Initiative (ADNI) is a longitudinal multisite study conducted at more than 60 clinical sites across the United States and Canada \citep{petersen2010}. Since its inception in 2004, ADNI has been continuously collecting data through multiple study phases, including ADNI1, ADNIGO, ADNI2, ADNI3, and the ongoing ADNI4 phase. The study enrolls cognitively normal older adults, individuals with mild cognitive impairment (MCI), and individuals with Alzheimer’s disease (AD). We restrict our analyses to participants diagnosed with MCI at baseline. The survival outcome of interest is the time from study entry to AD onset, which is interval-censored because cognitive status is assessed only at scheduled clinical visits. 

Our objective is to improve the prediction of AD risk among racial minority populations. To this end, we treat participants from non-White racial groups across all ADNI phases as the target cohort and use White participants as source cohorts. The covariates considered include age at enrollment (years), gender (male versus female), years of education, and the number of APOE-$\varepsilon4$ alleles (0, 1, or 2).

After removing participants with missing covariate information, a total of 152 subjects remain in the target cohort. Among them, 132 (86.84\%) are right-censored and 19 (12.50\%) are interval-censored, with a mean follow-up of 2.01 years and a maximum of 18.53 years. Since the adequacy of the Cox model for the target cohort has been validated using the supremum test of \citet{Xu2025checking}, we use the Cox model as the target model in both the single-source and multi-source analyses. 

In the single-source analysis, White participants from all ADNI phases are combined into a single source cohort. After applying the same data processing procedure as for the target cohort, the source cohort consists of 932 subjects. Among them, 580 (62.23\%) are right-censored and 330 (35.41\%) are interval-censored. The mean and maximum follow-up times are 3.92 years and 18.48 years, respectively. Table \ref{tab:cov1} of the Supplementary Material summarizes the clinical characteristics of the target and source cohorts and suggests only slight shifts in the covariate distributions.  

Similar to $\mathsection$\ref{subsec-single-sim}, we compare SPOT-IC with the target-only method, the combined method, Trans-Cox-Lasso, and Trans-Cox-MIC. We randomly split the target data into training and validation subsets at a $4:1$ ratio, and evaluate the prediction performance using IBS and negative log-likelihood (NLL) on the validation data, for which smaller values indicate better performance. The optimal logarithmic transformation models for the source data and the combined data are selected based on \AIC, yielding optimal values of $r=0.4$ and $r=0.1$, respectively. As shown in Table \ref{table:app1}, SPOT-IC substantially improves both prediction metrics relative to the target-only method. Moreover, it achieves near-best performance under both IBS and NLL among all competing methods.  
\begin{table}[htbp]
\tbl{Prediction results in the ADNI application with a single source}
{
\begin{tabular}{lccccc}
Metric & SPOT-IC & Target-only & Combined & Trans-Cox-Lasso & Trans-Cox-MIC \\
IBS & 0.103 (0.027) & 0.157 (0.052) & 0.101 (0.026) & 0.133 (0.044) & 0.125 (0.038) \\
NLL & 13.148 (4.726) & 19.692 (19.022) & 13.203 (4.106) & 13.314 (5.300) & 12.672 (4.807) \\
\end{tabular}
}
\label{table:app1}
\begin{tabnote}
Median (median absolute deviation) over 100 replicates are reported.
\end{tabnote}
\end{table}

\begin{figure}[ht]
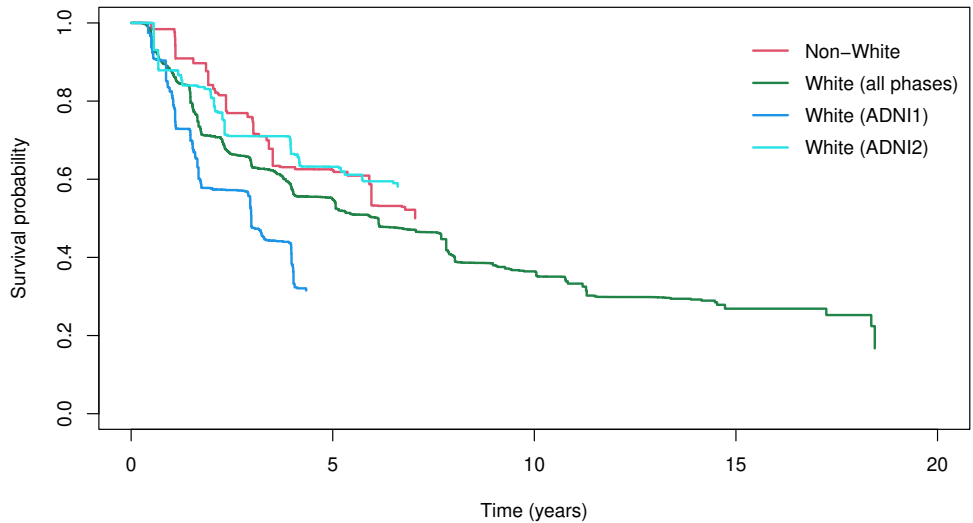

\figuresize{.6}
\figurebox{20pc}{25pc}{}[Tplot]
\caption{Turnbull estimators of the survival functions for the non-White (target) and White (source) cohorts in ADNI.}
\label{fig:turnbull1}
\end{figure}

In the multi-source analysis, we consider two source cohorts, which consist of White participants from ADNI1 and ADNI2, respectively. The ADNI1 source cohort contains 337 participants, of whom 176 (52.23\%) are right-censored and 140 (41.54\%) are interval-censored. Participants are followed for a mean of 2.72 years, up to a maximum of 6.76 years. The ADNI2 source cohort contains 433 participants, of whom 310 (71.59\%) are right-censored and 103 (23.79\%) are interval-censored. Participants are followed for a mean of 3.21 years, up to a maximum of 7.38 years. 
Figure~\ref{fig:turnbull1} displays the Turnbull estimators \citep{turnbull1976empirical} for the target and source cohorts.
The figure suggests that the ADNI2 source cohort has similar survival patterns to the target cohort and is thus likely to be informative, whereas the ADNI1 source cohort tends to experience AD onset earlier and is thus less informative. 
In addition, as shown in Table~\ref{tab:cov1}, both source cohorts share similar covariate distributions with the target cohort. 

% Figure \ref{fig:turnbull1} displays the Turnbull estimators for survival functions for these studies. The similarity in survival probabilities between the target and source studies suggests that borrowing information from source studies may improve the prediction results.

We compare SPOT-IC with the target-only method, Naive-SPOT-IC, TSDE, and MAE, as described in $\mathsection$\ref{subsec-multi-sim}. The evaluation procedure is the same as in the single-source analysis. For TSDE and MAE, Cox models are used for the source data because these two methods require the source and target models to share the same model structure. The prediction results are presented in Table \ref{table:app2}. SPOT-IC achieves the smallest IBS among all competing methods and also an NLL comparable to that of the best-performing method. These results further demonstrate the effectiveness of SPOT-IC in leveraging information from multiple heterogeneous source studies.
\begin{table}[htbp]
\tbl{Prediction results in the ADNI application with two sources}
{
\begin{tabular}{lccccc}
Metric & SPOT-IC & Target-only & Naive-SPOT-IC & MAE & TSDE \\
IBS & 0.109 (0.027) & 0.157 (0.052) & 0.121 (0.029) & 0.122 (0.042) & 0.122 (0.039) \\
NLL & 12.594 (4.417) & 19.692 (19.022) & 13.312 (4.993) & 12.447 (4.646) & 12.629 (4.904) \\
\end{tabular}
}
\label{table:app2}
\begin{tabnote}
Median (median absolute deviation) over 100 replicates are reported.
\end{tabnote}
\end{table}

% The results from both settings consistently suggest the practical utility of SPOT-IC for survival prediction compared to existing methods. 

\section{Discussion}\label{sec-disc}
% In this article, we propose a novel survival-probability-based transfer learning framework for improving prediction with interval-censored data. Compared with existing methods, our approach relies on a weaker similarity assumption, accommodates arbitrary source estimators, and does not require access to either the informativeness or the individual-level data of the source studies. These features make the method particularly attractive in practical applications. We show that the proposed estimator can achieve a faster convergence rate than the target-only estimator when at least one source estimator is sufficiently close to the true target survival function in $L_2$-distance. We further demonstrate the effectiveness of our approach in realistic settings through extensive numerical studies.

% This work proposes a novel survival-probability-based transfer learning framework for improving prediction with interval-censored data. Compared with existing methods, our approach relies on a weaker similarity assumption, accommodates arbitrary source estimators, and does not require access to either the informativeness or the individual-level data of the source studies. These features are particularly useful when external studies are only available through fitted prediction models or published risk predictions.

In $\mathsection$\ref{sec-app}, we treated White ADNI participants as the source cohorts, whose individual-level data are available, and used a common set of covariates between the target and source cohorts. This setup ensures a fair comparison with existing methods. Crucially, however, our SPOT-IC approach does not require access to individual-level source data or exact covariate alignment across studies. These features make SPOT-IC uniquely suited for external, privacy-sensitive sources or online risk calculators that restrict inputs to only simple covariates.

The current framework combines candidate estimators via Q-aggregation, where the aggregation weights are shared across all target subjects. This may be suboptimal when the target population itself is heterogeneous, for example, when it contains latent subgroups with distinct risk profiles. In such settings, it may be more appropriate to allow subgroup-specific or even subject-specific aggregation weights, so that greater weight is assigned to candidate estimators that are most relevant to a particular subgroup or individual.

Moreover, in the current framework, source information is essentially discarded once a source study is deemed noninformative for the target population. However, a biased source estimator may still become useful after appropriate calibration to the target distribution. This idea is closely related to prediction-powered inference \citep{angelopoulos2023prediction,angelopoulos2024ppiefficientpredictionpoweredinference,zou2026generalizedpredictionpoweredinferenceapplication}, which improves statistical efficiency by calibrating and debiasing imperfect predictions. In survival analysis, AI-generated survival predictions are often trained on large external cohorts and may therefore be systematically biased for a target population. Developing transfer learning methods that can calibrate and debias such predictions is a promising direction for future research.

\section*{Acknowledgment}
This research was supported by the Hong Kong Research Grant Council grant 27303624 and research funds from the University of Hong Kong.

% \section*{Disclosure Statement}
% The authors report there are no competing interests to declare.

% \section*{Data Availability Statement}
% The data that support the findings of this study are openly available at \url{https://adni.loni.usc.edu/}.

\section*{Supplementary material}
We provide the proofs of the theorems in $\mathsection$\ref{sec-theory}, lemmas and their proofs, and an additional table referenced in $\mathsection$\ref{sec-app}.

\appendixone
\section*{Appendix}
\subsection*{Details of EM algorithm}
This appendix provides technical details of the EM algorithm described in $\mathsection$\ref{sec-meth}. The complete-data weighted log-likelihood is
\[
\begin{aligned}
    n^{-1} &\sum_{i=1}^n \left(\sum_{l:t_l \le R^*_i} \left[W_{il}\log\left\{z_i \lambda_l \exp{(\bb^{\T} \X_{il})}\right\}-z_i\lambda_l\exp{(\bb^{\T}\X_{il})}-\log(W_{il}!) \right] + \log f(z_i) \right) \\
    & + \xi_n (mJ)^{-1} \sum_{i=1}^{m} \sum_{j=1}^{J} \sum_{l:t_l \le \tilde{Y}_i} \left[\tilde{W}_{ijl}\log\left\{\tilde{z}_i \lambda_l \exp{(\bb^{\T} \tilde{\X}_{il})}\right\}-\tilde{z}_i\lambda_l\exp{(\bb^{\T} \tilde{\X}_{il})}-\log(\tilde{W}_{ijl}!) \right]\\
    &+ \xi_n m^{-1} \sum_{i=1}^{m} \log f(\tilde{z}_i),
\end{aligned}
\]
where $R_i^* = L_i I(R_i=\infty)+R_iI(R_i<\infty)$.

In the M-step, define
\begin{equation*}
    s^{(0)}(t;\bb) = n^{-1}\sum_{i=1}^n I(t\le R_i^*)\hat{E}(z_i)e^{\bb^{\T}\X_i(t)}+\xi_nm^{-1}\sum_{i=1}^m I(t\le \tilde{Y}_i)\hat{E}(\tilde{z}_i)e^{\bb^{\T}\tilde{\X}_i(t)},
\end{equation*}
\begin{equation*}
    s^{(1)}(t;\bb) = n^{-1}\sum_{i=1}^n I(t\le R_i^*)\hat{E}(z_i)e^{\bb^{\T}\X_i(t)}\X_i(t)+\xi_nm^{-1}\sum_{i=1}^m I(t\le \tilde{Y}_i)\hat{E}(\tilde{z}_i)e^{\bb^{\T}\tilde{\X}_i(t)}\tilde{\X}_i(t).
\end{equation*}
We first update $\lambda_l$ $(l=1,\dots,L)$ by
\[
    \lambda_l = \frac{n^{-1}\sum_{i=1}^nI(t_l\le R_i^*)\hat{E}(W_{il})+\xi_nm^{-1}\sum_{i=1}^m I(t_l\le \tilde{Y}_i)\hat{E}(\tilde{W}_{il})}{s^{(0)}(t_l;\bb)},
\]
where $\tilde{W}_{il} = J^{-1}\sum_{j=1}^J \tilde{W}_{ijl}$.
Then we 
% incorporate \eqref{eq:lam update} into the conditional expectation of \eqref{eq:loglik} and 
update $\bb$ by solving the following equation using the one-step Newton--Raphson method:
\begin{equation*}
    n^{-1}\sum_{i=1}^n \sum_{l:t_l \le R_i^*} \hat{E}(W_{il})\left\{ \X_{il}-\frac{s^{(1)}(t_l;\bb)}{s^{(0)}(t_l;\bb)}\right\} +\xi_n m^{-1} \sum_{i=1}^m \sum_{l:t_l \le\tilde{Y}_i} \hat{E}(\tilde{W}_{il})\left\{\tilde{\X}_{il}-\frac{s^{(1)}(t_l;\bb)}{s^{(0)}(t_l;\bb)}\right\}=\mbf{0}.
\end{equation*}

In the E-step, we evaluate $\hat{E}(W_{il})$, $\hat{E}(\tilde{W}_{il})$, $\hat{E}(z_i)$ and $\hat{E}(\tilde z_i)$, where $\hat{E}(\cdot)$ denotes the conditional expectation given the observed data. For each $i \in \{1,\dots,n\}$, define $Q_{i1} = \sum_{l:t_l\le L_i} \lambda_l\exp(\bb^{\T}\X_{il})$ and $Q_{i2} = \sum_{l:t_l\le R_i} \lambda_l \exp(\bb^{\T}\X_{il})$. The posterior density for $z_i$ is proportional to $\{\exp(-z_iQ_{i1})-\exp(-z_iQ_{i2})\}f(z_i)$,
which yields
\begin{equation*}
\hat{E}(z_i)=\frac{\exp\{-G(Q_{i1})\}G'(Q_{i1})-\exp\{-G(Q_{i2})\}G'(Q_{i2})}{\exp\{-G(Q_{i1})\}-\exp\{-G(Q_{i2})\}}.
\end{equation*}
Similarly, the posterior density for $\tilde{z}_i$ is proportional to
\begin{equation*}
\left\{ \exp\left( - \sum_{l : t_l \leq \tilde{Y}_i} \tilde{z}_i \lambda_l e^{{\bb}^{\T} \tilde{\X}_{il}} \right) \right\}^{\check{S}(\tilde{Y}_i | \tilde{\X}_i)}
\times \left\{ 1 - \exp\left( - \sum_{l : t_l \leq \tilde{Y}_i} \tilde{z}_i \lambda_l e^{{\bb}^{\T} \tilde{\X}_{il}} \right) \right\}^{1 - \check{S}(\tilde{Y}_i | \tilde{\X}_i)}
f(\tilde{z}_i).
\end{equation*}
All integrals over $z_i$ and $\tilde{z}_i$ can be approximated using Gauss--Laguerre quadrature.

For the Poisson variables, it is easy to see that $\hat{E}(W_{il}) = 0$ for $t_l \le L_i$,
and for $L_i <t_l \le R_i$ with $R_i < \infty$,
\begin{equation*}
\begin{aligned}
    \hat{E}(W_{il}) & = \hat{E}_{z_i}\left\{\hat{E}\left(W_{il}|z_i\right)\right\} = \hat{E}_{z_i}\left\{\frac{z_i\lambda_l\exp(\bb^{\T}\X_{il})}{1-\exp\left\{-z_i\left(Q_{i2}-Q_{i1}\right)\right\}}\right\}\\[2mm]
    & = \lambda_l\exp(\bb^{\T}\X_{il}) \\
    & \quad \times \frac{\int_{z_i}z_i[1 - \exp\{-z_i(Q_{i2} - Q_{i1})\}]^{-1}\left\{ \exp(-z_i Q_{i1}) - \exp(-z_i Q_{i2}) \right\} f(z_i)dz_i}{\exp\{-G(Q_{i1})\} - \exp\{-G(Q_{i2})\}}.
\end{aligned}
\end{equation*}
In addition,
the conditional expectation of $W_{ijl}$ given the observed data and $\tilde{z}_i$ is
\begin{equation*}
\begin{aligned}
    \hat{E}(\tilde{W}_{ijl}|\tilde{z}_i)
    & = \frac{\delta_{ij}\tilde{z}_i\lambda_l\exp(\bb^{\T}\tilde{\X}_{il})}{1-\exp\left\{-\sum_{l':t_{l'}\le \tilde{Y}_i} \tilde{z}_i\lambda_{l'} \exp(\bb^{\T}\tilde{\X}_{il'}) \right\}}.
\end{aligned}
\end{equation*}
Thus, given the fact that $\lim_{J\to\infty} J^{-1} \sum_{j=1}^{J} \delta_{ij} = 1 - \check{S}(\tilde{Y}_i | \tilde{\X}_i)$, the conditional expectation of $\tilde{W}_{il}$ given the observed data and $\tilde{z}_i$ is
\begin{equation}\label{eq:post-tilde-W}
\begin{aligned}
    \hat{E}(\tilde{W}_{il}|\tilde{z}_i) = \frac{\{1 - \check{S}(\tilde{Y}_i | \tilde{\X}_i)\}\tilde{z}_i\lambda_l\exp(\bb^{\T}\tilde{\X}_{il})}{1 - \exp\left\{-\sum_{l':t_{l'}\le \tilde{Y}_i} \tilde{z}_i\lambda_{l'} \exp(\bb^{\T}\tilde{\X}_{il'})\right\}}.
\end{aligned}
\end{equation}

The initial value for $\bb$ is set to be $\mbf 0$, and $\lambda_l$  is initialized as $1/L$. We iterate between the E-step and M-step until both the $L_2$-difference in the $\bb$ estimates and the sum of the absolute differences in the $\lambda_l$ estimates between successive iterations are smaller than the prespecified threshold, such as $10^{-3}$.

\bibliographystyle{biometrika}
\bibliography{bibliography}

\end{document}